\documentclass[a4paper,12pt]{article}
\pdfoutput=1
\usepackage[a4paper, bindingoffset=0.2in, left=1in, right=1in, top=1in, bottom=1in, footskip=.25in]{geometry}

\usepackage{times}
\usepackage{amsmath,amssymb}
\usepackage{graphicx,afterpage}
\usepackage[final]{pdfpages}
\usepackage{authblk}
\usepackage[nottoc]{tocbibind}
\usepackage{setspace}
\usepackage{bbm}
\usepackage{hyperref}
\usepackage{xcolor}
\hypersetup{
  colorlinks   = true, 
  urlcolor     = red, 
  linkcolor    = blue, 
  citecolor   = blue 
}
\usepackage[labelsep=period, labelfont=sc]{caption}
\usepackage{tabularx,ragged2e,booktabs}
\usepackage{sectsty}

\usepackage{caption}
\usepackage{lineno}

\usepackage{enumitem}
\title{Statistical Criticality arises in Most Informative Representations}

\author[1,2,3]{\small Ryan John Cubero}
\author[4,5,6,7]{Junghyo Jo} 
\author[2,8]{\small Matteo Marsili} 
\author[1]{\small Yasser Roudi}
\author[2,5,6]{Juyong Song}

\affil[1]{\footnotesize Kavli Institute for Systems Neuroscience and Centre for Neural Computation, Norwegian University of Science and Technology (NTNU), Olav Kyrres gate 9, 7030 Trondheim, Norway}
\affil[2]{\footnotesize The Abdus Salam International Center for Theoretical Physics, Strada Costiera 11, 34151 Trieste, Italy}
\affil[3]{\footnotesize Scuola Internazionale Superiore di Studi Avanzati, Via Bonomea 265, 34136 Trieste, Italy}
\affil[4]{Department of Statistics, Keimyung University, Daegu 42601, Republic of Korea}
\affil[5]{Asia Pacific Center for Theoretical Physics, Pohang, Gyeongbuk 37673, Korea}
\affil[6]{Department of Physics, Pohang University of Science and Technology, Pohang, Gyeongbuk 37673, Korea}
\affil[7]{School of Computational Sciences, Korea Institute for Advanced Study, Seoul 02455, Korea}
\affil[8]{Istituto Nazionale di Fisica Nucleare (INFN), Sezione di Trieste, Italy}

\date{17 June 2019}

\begin{document}
\maketitle

\begin{abstract}
We show that {\em statistical criticality}, i.e. the occurrence of power law frequency distributions, arises in samples that are maximally informative about the underlying generating process. In order to reach this conclusion, 
we first identify the frequency with which different outcomes occur in a sample, as the variable carrying useful information on the generative process. The entropy of the frequency, that we call {\em relevance}, provides an upper bound to the number of {informative} bits.  
This differs from the entropy of the data, that we take as a measure of {\em  resolution}. 
Samples that maximise relevance at a given resolution -- that we call {\em maximally informative samples} -- exhibit statistical criticality. In particular, Zipf's law arises at the optimal trade-off between resolution (i.e. compression) and relevance. As a byproduct, we derive a bound of the maximal number of parameters that can be estimated from a dataset, in the absence of prior knowledge on the generative model. 

Furthermore, we relate criticality to the statistical properties of the representation of the data generating process. We show that, as a consequence of the concentration property of the Asymptotic Equipartition Property, representations that are maximally informative about the data generating process are characterised by an exponential distribution of energy levels. This arises from a principle of minimal entropy, that is conjugate of the maximum entropy principle in statistical mechanics. This explains why statistical criticality requires no parameter fine tuning in maximally informative samples.

\end{abstract}

When data are generated as independent draws from a parametric distribution, one can draw a sharp distinction between noise and {\em useful} information, that part of the data that can be used to estimate the generative model. Useful information is concentrated in sufficient statistics, which are those variables whose empirical value suffices to fully estimate the model's parameter \cite{statistics}.  
The first aim of this paper is to draw the same distinction in the case where the model is not known. In this case, we show that the information on the generative model is contained in the distribution of frequencies, i.e. the fraction of times different outcomes occur in the dataset. Indeed, frequencies provide a  {\em minimally sufficient representation} of the sample, analogous to that of sufficient statistics, since they encode all relevant information on the generative process. Therefore, the amount of information that the sample contains on the generative process is given by the entropy of the frequency distribution that, following Ref. \cite{HM}, we call {\em relevance}. The relevance is only part of the total information contained in the sample. The total information, on the other hand, is quantified by the entropy of the distribution of outcomes and, as argued in Ref. \cite{HM}, is a measure of {\em resolution}\footnote{For example, stocks in the financial market can be classified by their SIC (Standard Industrial Classification) code using different number of digits, gene sequences can be defined in terms of the sequence of the bases or in terms of the sequence of amino acids they code for, etc.}. 
The relevance provides an upper bound on the number of parameters that can be inferred from a sample, in the absence of prior information on the model. This characterisation also allows us to define {\em maximally informative samples}, which are those that maximise the relevance at a fixed resolution. 

As shown in Refs. \cite{HM,MMR}, maximally informative samples in the under sampling regime exhibit {\em statistical criticality} \cite{Mora,statcrit}. This implies that the number of outcomes that occur $k$ times in the sample behaves as $m_k\sim k^{-\mu-1}$. Here, the exponent $\mu$ encodes the trade-off between resolution and relevance: a decrease of one bit in resolution affords an increase of $\mu$ bits in relevance. Hence, the case $\mu=1$, which corresponds to the celebrated Zipf's law\footnote{Zipf's law corresponds to the statement that the $r^{\rm th}$ most frequent outcome occurs a number of times which is inversely proportional to $r$. Hence, Zipf's law manifests as a straight line in a log-log rank-frequency plot, with slope equal to $-1/\mu=-1$.} \cite{Zipf}, encodes the optimal trade-off, since further decrease in resolution delivers an increase in relevance that does not compensate for the information loss. 

The second aim of this paper is to characterise the properties of the generating process itself, under the assumption that it provides a {\em maximally informative representations} of an underlying complex system. Our main argument is that if a maximally informative representation extracts the optimal features from the data, then, conditional to it, the generated data should appear as noise. Then, the Asymptotic Equipartition Property \cite{CoverThomas} ensures that the logarithm of the probability of a typical data point conditional on the representation should concentrate, i.e. it should have a narrow distribution. This identifies the log-probability -- i.e. minus the {\em energy}, in a statistical mechanics analogy -- as the natural variable in the representations. Our central result is that maximally informative representations are characterised by an exponential distribution of energy levels. 
Projected onto a finite dataset, this again identifies the frequency as the relevant variable. Non-trivial statistical dependencies in the data reveal themselves in a wide variation in the energy and this, in turn,  is a necessary condition for the occurrence of Zipf's law, as shown in \cite{MSN,ACL}. In general, we confirm that power law distributions in the frequency emerge as a natural consequence of most informative representations at different levels of resolution (or compression). 

Statistical criticality has attracted considerable attention in statistical physics, because it is reminiscent of critical phenomena, but it arises without the need to fine tune parameters to special points (see e.g. \cite{Mora, statcrit,Hidalgo}). We relate this apparent puzzle to the fact that, while statistical physics is based on the {\em maximisation} of the entropy over the distributions on the micro-states, most informative representations arise from a different optimisation problem: Energy levels correspond to information costs, which are defined in terms of the probabilities of the micro-states. The (analog of the Boltzmann) entropy provides an intrinsic measure of the noise, i.e. of the degeneracy of energy levels that the representation is not able to resolve. Maximally informative representations are those that {\em minimise} this entropy, with respect to the energy density, at a given coding cost (average energy). 
Broad distributions occur in general, without the need of fine tuning, in the solution to the latter problem. This suggests that we expect statistical criticality whenever the data is expressed in terms of most informative representations. When applied, for example, to cities \cite{city} or to the distribution of firms by SIC codes \cite{dissecting}, the occurrence of broad distributions suggests that city names and SIC codes are relevant labels, because they provide a highly informative representation. It is also interesting to note that systems that are meant to encode efficient representations follow Zipf's law. This is the case, for example, for the frequency of words in language \cite{Zipf}, the antibody binding site sequences in the immune system \cite{Immune,Mora5405} and spike patterns of population of neurons \cite{retina}. We regard these examples as independent checks, which support our main claim that the occurrence of statistical criticality in a sample can be taken as a {\em certificate of the efficiency} of the representation used. 

It is important to remark that the term {\em efficient representation} has often been used with respect to some input stimuli \cite{Chalk186} or in terms of predictive information \cite{IlyaBialek,EffRepBialek}. Here, it refers to the statistical properties of the representation itself, independently of the nature of the input, as long as this is non-trivial.
Just like entropy measures information content irrespective of what the information is about, we show that a quantitative measure of relevance is possible without reference to what relevance refers to. As a corollary, maximally informative representations can be defined, without explicit reference to what is represented. This allows one to disentangle the problem of efficiently encoding an input to that of understanding what the representation describes, that are intertwined in approaches to efficient coding based on input-output relations \cite{Chalk186}.
As such, we believe that our results may provide a guiding principle to extract relevant variables from high dimensional data (see e.g. \cite{GMF,CMR}), or to shed light on the principles underlying deep learning (see e.g. \cite{Hennig,SMJ}). 


Finally, we show that maximally informative samples can be derived from the Information Bottleneck (IB) \cite{IB}  approach, when the frequency is taken as the output variable. This turns the IB into an unsupervised learning approach to learn the underlying generative process.

\section{Minimally sufficient representations}
\label{MSR}

Consider a sample of $N$ data points $\hat s=(s_1,\ldots, s_N)$, each being drawn from some alphabet $\mathcal{S}$. The only information we shall consider is the one contained in the sample, as in an {\em unsupervised} learning setting. 
We assume that $s_i$ are outcomes of $N$ independent observations, so the order of the data points is irrelevant. Mathematically, this is equivalent to each $s_i$ being independent and identically distributed draws from an unknown distribution $p(s)$, that we shall call the (unknown) generative model. 
In order to keep our discussion as general as possible, we make no assumption on $\mathcal{S}$ which may even be unknown in advance (e.g. when sampling species from an unexplored ecosystem), or on the structure of the outcomes (e.g. $s_i$ could be words, protein sequences or bit strings, etc). In brief, we shall consider $s_i$ as abstract labels. We shall postpone the discussion on how further information affects our results to later sections (see Section 3).

The information content of the sample can be quantified in the number of bits needed to represent one of the outcomes, which is given by the entropy
\begin{equation}
\label{totH}
\hat H[s]=-\sum_{s\in\mathcal{S}}\hat p_s\log_2 \hat p_s,\qquad\hat p_s=\frac{k_s}{N}
\end{equation}
where $\hat p_s$ is the empirical distribution and $k_s$ is the number of points in the sample with $s_i=s$. 
We stress that we refer to the entropy in Eq. (\ref{totH}) as a quantitative measure of description length rather than as an estimate of the true entropy of an underlying distribution\footnote{For a given distribution $p(s)$, Eq. (\ref{totH}) is known to provide a (positively) biased estimate of the true entropy $H[s]=-\sum_s p(s)\log p(s)$, for a finite sample, because of the convexity of the logarithm \cite{miller1955note,entropy_est}.}. Henceforth, we shall use the $\hat{~}$ to denote entropies measured from empirical distributions, as in Eq. (\ref{totH}).
Some of the $\hat H[s]$ bits convey useful information on the generative process, some bits are just noise. 
The precise definition of noise in this present context relies on the maximum entropy principle, which encodes a state of maximal ignorance \cite{Jaynes}. Therefore a random variable is noise if it has a maximum entropy distribution. 

Our aim is to provide an upper bound to the number of {\em useful} bits. 
In order to do this, we will search for a set of {\em hidden features} such that {\em i)} conditional on these features, the data is as random as possible (in the sense of maximal entropy) and can be considered as noise, and that {\em ii)} provide the most concise representation (in terms of description length) of useful information or, equivalently, that noise accounts for as much as possible of the sample's information content in Eq. (\ref{totH}). 
Features defined in this way provide what we call a {\em minimally sufficient representation}, in the sense that they carry the maximal possible amount of useful information on the generative process. In this sense, such features provide the most concise representation of the useful information.

Let $\hat h=(h_1,\ldots,h_N)$ be a set of variables -- the features -- with $h_i\in \mathcal{H}$ taking values in a finite set. With the introduction of these variables, the dataset is augmented to $\hat d=(\hat s,\hat h)$ in such a way that the total information content becomes
\[
\hat H[s,h]=\hat H[s]+\hat H[h|s].
\]
As a first requirement, we demand that the features do not introduce additional information. This means that $\hat H[h|s]=0$ or equivalently, that the features $h_i=h(s_i)$ are function of the data $s_i$. This allows us to separate the total information content into a part that depends on $h$ and a part that depends on the data $s$ conditional on $h$:
\begin{equation}
\label{eq:infosplit}
\hat H[s]=\hat H[h]+\hat H[s|h].
\end{equation}
Our second requirement is that $\hat H[s|h]$ accounts only for noise in the sample. Put differently, the subsample of points $s_i$ is such that $h(s_i)=h$ should be consistent with a state of maximal ignorance, for all $h$. This means that 
the distribution $\hat p(s|h)$ over the $s$ for which $h(s)=h$ should be a distribution of maximal entropy\footnote{Let $\mathcal{S}_h=\{s: h(s)=h\}$ be the set of outcomes $s$ such that $h(s)=h$ and let $n_{h}=|\mathcal{S}_h|$ be the number of such outcomes. Then, the maximum entropy distribution is $\hat p(s|h)=1/n_h$ for all $s\in \mathcal{S}_h$ and $\hat p(s|h)=0$ otherwise, and its entropy is $\log_2 n_h$. } \cite{Jaynes}, i.e.
\begin{equation}
\label{maxent}
\hat p(s|h)=\hat p(s'|h) \qquad  \forall s,s'~\hbox{such that}~h(s)=h(s')=h.
\end{equation}
Since $\hat p(s)=\hat p(s|h(s))\hat p(h(s))=k_s/N$, this in turn implies that outcomes $s$ and $s'$ that are assigned the same feature, should have the same frequency, i.e. $k_s=k_{s'}$ whenever $h(s)=h(s')$, and that $h(s)\neq h(s')$ if $k_s\neq k_{s'}$. As a result of this, $k_s$ can be expressed as a function of $h(s)$. The data processing inequality \cite{CoverThomas} then implies that, 
\begin{eqnarray}
\hat H[h]\ge\hat H[k]. \label{ineq}
\end{eqnarray}
Note that the choice $h(s)=s$ would trivially satisfy the requirement in Eq. (\ref{maxent}), but with $\hat H[s|h]=0$. 
Among all function $h(s)$ consistent with Eq. (\ref{maxent}), the minimally sufficient ones are those for which $\hat H[s|h]$ is as large as possible, or equivalently, those for which $\hat H[h]$ is as small as possible. 
Minimally sufficient representations are those that saturate the inequality in Eq. (\ref{ineq}), i.e. those where $h(s)=g(k_s)$ is a monotonous function of $k_s$, or without loss of generality simply, $h(s)=k_s$.

This leads us to the following:

\paragraph{Proposition} {\em The frequency $k_s$ provides a minimally sufficient representation of the sample $\hat s$ in the sense that:

\noindent
{\em i)} the total information content of a sample $\hat s$ can be divided as
\begin{equation}
\label{decamp}
\hat H[s]=\hat H[k]+\hat H[s|k]
\end{equation}
where
\begin{equation}
\label{Hk}
\hat H[k]=-\sum_k \frac{k m_k}{N}\log_2 \frac{k m_k}{N},
\end{equation}
with $m_k$ being the number of outcomes $s$ for which $k_s=k$, and
\begin{equation}
\label{Hsk}
\hat H[s|k]=\sum_k \frac{k m_k}{N}\log_2 m_k.
\end{equation}
\noindent
{\em ii)} In the absence of prior information, 
$\hat H[k]$ is the maximal number of bits (per data point) that can be used to estimate the underlying generative process and $\hat H[s|k]$ is a measure of noise.}
\vskip 0.5cm

In hindsight, this result is self-evident because, in the absence of prior information, the frequency $k_s$ with which different outcomes $s$ occur is the only statistics that can distinguish them. In order to illustrate this point, consider two outcomes $s$ and $s'$ that occur the same number of times $k_s=k_{s'}=k$. The distinction between outcomes $s$ and $s'$ is based on some pre-defined classification criteria\footnote{For example, the words "horse, coyote, house, castle" can be classified in two groups as (house, castle) and (horse, coyote), according to meaning, or (horse, house) and (castle, coyote) according to word length or of the occurrence of the same letters.}. Any other classification that distinguishes these $2k$ sample points into two classes of equal size $k$ would result in a sample with exactly the same statistics. The data contains no information that can distinguish the pre-defined classification from any other classification that yields the same class sizes. When $m_k$ outcomes are observed $k$ times, the number of classifications that are consistent with the data, is given by
\begin{equation}
\prod_k\frac{(km_k)!}{(k!)^{m_k}}\simeq e^{N \hat H[s|k]}=e^{N (\hat H[s]-\hat H[k])},
\end{equation}
where we used Stirling's approximation $n!\simeq n^ne^{-n}$ (assuming $km_k\gg 1$). This shows that $\hat H[s|k]$ is a measure of ambiguity of the representation used, because it captures the residual degeneracy on possible classification criteria that the data is not able to remove. Notice that in the well sampled regime, $N/|\mathcal{S}|\gg 1$, all states are expected to be sampled a different number of times and hence  $m_k\le 1$ for all $k$, which implies $\hat H[s|k]=0$.


\section{Resolution, relevance and maximally informative samples}
\label{maxinfosamples}

In typical cases, the dataset $\hat s$ encodes a description of a set $\hat x$ of complex objects at a given level of detail (e.g. proteins, texts, organisms, firms, etc). To fix ideas, we can think of such complex objects as a high dimensional vector $\vec x$, whose components provide a detailed description of all the characteristics of the objects, and to $s=s(\vec x)$ as a function taking values in a discrete set of labels $\mathcal{S}$. For the same data $\hat x=(\vec x_1,\ldots,\vec x_N)$, the variables $s_i=s(\vec x_i)$ can be chosen in different ways, i.e. with different levels of detail\footnote{We neglect the case where $\vec x_i=\vec x_j$ for some $i\neq j$, which is unlikely when the dimensionality of $\vec x$ is very large.}. Depending on this, the coding cost $\hat H[s]$ in Eq. (\ref{totH}) can take different values. At a sufficiently fine level of detail, each point is identified by different labels ($s(\vec x_i)\neq s(\vec x_j)$ for all $i\neq j$), which means that $k_s=0,1$ for all $s$ and hence, $\hat H[s]=\log N$. Any finer level of detail\footnote{For example, given a list of tags, $s(\vec x)$ could specify which tags belong to image $\vec x$. Adding one more tag $\tau(\vec x)$ to the list, i.e. $s'(\vec x)=(s(\vec x),\tau(\vec x))$, corresponds to increasing the level of detail.} corresponds to a mere relabelling of the objects, and hence, the information content $\hat H[s]=\log N$ remains the same. At the other extreme, when the description is so coarse that all objects correspond to the same label ($s(\vec x_i)=s_0$ for all $i$), the coding cost vanishes $\hat H[s]=0$.

Following Ref. \cite{HM}, we shall call $\hat H[s]$ {\em resolution}, since it quantifies in bits the level of detail of the description $s(\vec x)$ in the sample $\hat s$. 

An intermediate value of $\hat H[s]$ may correspond to different possible representations $s(\vec x)$. For each of these representations, the entropy $\hat H[k]$ is an intrinsic property of the sample in that level of detail and, as argued above, it quantifies the amount of information that the sample contains on the generative model. For this reason, we shall follow Ref. \cite{HM} and call $\hat H[k]$ {\em relevance}. From Eq. (\ref{Hk}), it is clear that $\hat H[k]=0$ in both the extreme cases discussed above. Indeed, when $\hat H[s]=\log N$, $m_k=0$ for all $k>1$ and $m_1=N$, whereas when $\hat H[s]=0$, $m_k=0$ for all $k<N$ and $m_N=1$. 

The trade-off between resolution and relevance 
can be visualised in a plot of $\hat H[k]$ as a function of $\hat H[s]$, as the latter varies between $0$ and $\log N$.
In between the extreme cases $\hat H[s]=0$ and $\hat H[s]=\log N$, $\hat H[k]$ follows a bell shaped curve that is upper bounded by the line $\hat H[k]=\hat H[s]$, that is attained when $m_k=0$ or $1$ for all values of $k$. This is because $k_s$ is a function of $s$ and the data processing inequality imposes that it cannot contain more information than $s$ itself. The region on the right of the maximum is what we shall refer to as the under-sampling regime.

\begin{figure}[ht]
\centering
\includegraphics[width=0.8\textwidth]{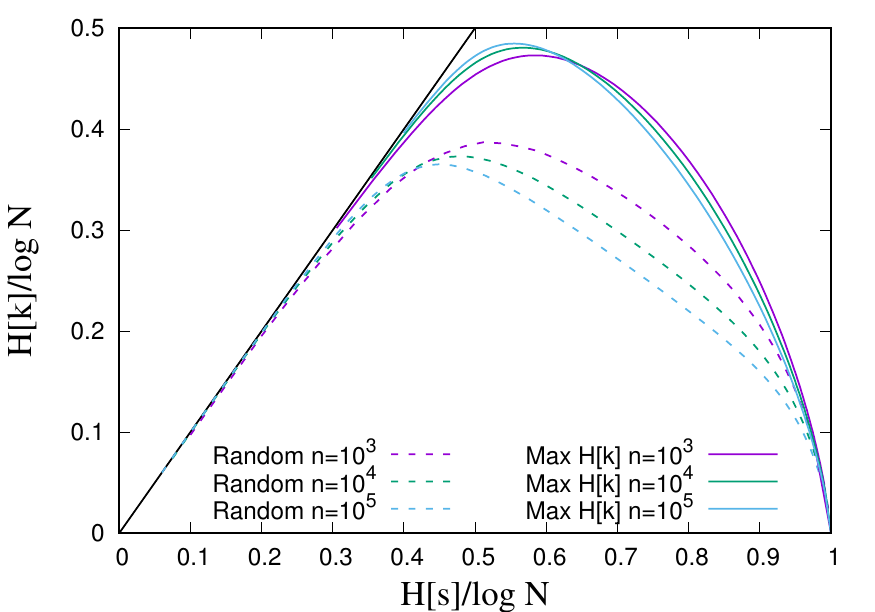}
\caption{\label{fig:rand_vs_opt} Plot of $\hat H[k]$ versus $\hat H[s]$ for maximally informative samples (solid lines) and for structure-less samples (dashed lines), for $N=10^3, 10^4$ and $10^5$. For maximally informative samples, we report the lower bound derived in Ref. \cite{HM}. For structure-less samples, we compute the averages of $\hat H[s]$ and $\hat H[k]$ over $10^7$ realisations of random distributions of $N$ balls in $L$ boxes, with $L$ varying from $2$ to $10^7$. Here, each box corresponds to one state $s=1,\ldots,L$ and $k_s$ is the number of balls in box $s$. The full black line represents the limit $\hat H[k]=\hat H[s]$ which is obtained when $m_k \le 1$ for all $k$. Points above this line are ruled out by the data processing inequality.}
\end{figure}

The curve depends on the structure of $\vec x$ and on how it is captured by the representation adopted at different resolutions. For structure-less samples $\hat x$, we expect that the statistics of $\hat s$ is equivalent to drawing $N$ balls at random in $L$ boxes\footnote{This is exactly true if $\vec x\in [0,1]^D$ is a vector of $D\gg 1$ independent uniformly distributed components $x^{(a)}$. Then $s(\vec x)$ can be taken as the labels of $L$ non-overlapping subsets of $[0,1]^D$ of volume $1/L$. A similar construction can be done for the case where components $x^{(a)}$ are independently drawn with pdf $f_a(x)$, because the variables $z^{(a)}=\int_{-\infty}^{x^{(a)}} f_a(x)dx$ would be uniformly distributed in $[0,1]^D$.}. Fig. \ref{fig:rand_vs_opt} shows the behaviour of $\hat H[k]$ as a function of $\hat H[s]$ for structure-less samples (dashed lines) as $L$ is varied. 

One can compare this curve with the one obtained for samples of a given size $N$ and resolution $\hat H[s]$, with maximal relevance $\hat H[k]$. These are the samples that we call {\em most informative samples}, and they can be derived from the solution of the problem 
\begin{equation}
\label{maxHk}
\max_{m_k}\left[\hat H[k]+\mu\hat H[s]+ \lambda\sum_{k} km_k\right]
\end{equation}
with $\hat H[s]=-\sum_k\frac{km_k}{N}\log \frac{k}{N}$ and $\hat H[k]$  given by Eq. (\ref{Hk}). In Eq. (\ref{maxHk}), $\mu$ and $\lambda$ are Lagrange multipliers enforcing the constraints on $\hat H[s]$ and $N$.
Given that $m_k$ is an integer variable, the maximisation cannot be performed analytically. Fig.  \ref{fig:rand_vs_opt} reports a lower bound  (full lines) obtained in \cite{HM} by taking $m_k$ as Poisson variables with mean $\bar m_k$ and maximising the expected value of $\hat H[k]$ over the latter, at fixed expected values of $\hat H[s]$ and sample size $N$. As the plot shows, the difference between the (lower-bound of the) maximal achievable curve and the one that refers to random samples increases as $N$ increases. As shown in \cite{MMR,HM}, and by a direct solution of Eq. (\ref{maxHk}) with real $m_k\ge 0$, the maximally informative samples in the under-sampling regime have a characteristic power law frequency distribution
\begin{equation}
\label{powlaw}
m_k\sim k^{-\mu-1}.
\end{equation}
Eq. (\ref{powlaw}) shows that statistical criticality, i.e. the occurrence of power law frequency distributions in samples, is a signature of maximally informative samples.

In the rightmost part of Fig.  \ref{fig:rand_vs_opt}, $\hat H[k]$ decreases as $\hat H[s]$ increases with a slope $-\mu$, which is the Lagrange multiplier enforcing the constraint on $\hat H[s]$ in Eq. (\ref{maxHk}). Because of this constraint, a decrease of $\Delta$ bits in $\hat H[s]$ grants an increase of $\mu\Delta$ bits in $\hat H[k]$. Hence, $\mu$ quantifies the trade-off between resolution ($\hat H[s]$) and relevance ($\hat H[k]$), as observed in Ref. \cite{SMJ}. The point $\mu=1$ marks the limit beyond which further reduction in the resolution implies losses in accuracy. For this reason, this point separates the region of lossless compression ($\mu\ge 1$) from the region of lossy compression ($\mu<1$) and is the one for which $\hat H[s]+\hat H[k]$ is maximal. The parameter $\mu$ is also the exponent of the frequency distribution in Eq. (\ref{powlaw}). Therefore, the point $\mu=1$ corresponds to Zipf's law, suggesting that the occurrence of Zipf's law is a signature of an efficient representation at the optimal trade-off between resolution and relevance. 

\section{Relation with parametric models} 

The features $\hat k$ play the same role of minimally sufficient statistics when the data is generated from a known parametric model $p(s|\theta)$. Minimal sufficient statistics are those combination $T(\hat s)$ of the data that contain all information about the model parameters $\theta$ \cite{statistics}. In other words, the mutual information between $\hat s$ and $\theta$ equals the mutual information between $T(\hat s)$ and $\theta$. By the Neyman-Fisher factorisation theorem \cite{statistics}, the probability of samples $\hat s$ conditional on $T(\hat s)=t$ does not depend on $\theta$, i.e. it only encodes noise. Notice that $p(\hat s|T(\hat s)=t)$ is independent of $\hat s$, i.e. it is a maximum entropy distribution, consistent with our definition of noise. Analogously to minimal sufficient statistics $T(\hat s)$, the frequencies $\hat k$ provide the same sharp separation between useful information and noise, in a model-free setting. 

Of course, the knowledge that $\hat s$ comes from a given distribution $p(s|\theta)$ changes considerably the picture\footnote{Even the presence of a structure in the data points provides useful information beyond what is assumed in our general setting. For example, if $s=(\sigma_1,\ldots,\sigma_n)$ is a configuration of $n$ spins $\sigma_i=\pm 1$ (or bits $\sigma_i=0,1$) then one expects that configurations that differ by the value of few $\sigma_i$'s should have similar probabilities. Such expectation favours models with low order interactions.}. In that case, the information that the sample $\hat s$ contains on the generative model can be quantified in the Kullback-Leibler divergence between the posterior distribution $p(\theta|\hat s)$ and the prior $p_0(\theta)$. For $N$ large, a straightforward calculation yields 
\begin{equation}
\label{eq:Dkl}
D_{KL}\left(p(\theta|\hat s)||p_0(\theta)\right)\simeq \frac{d}{2}\log \frac{N}{2\pi e}+\frac{1}{2}\log{\rm det} \hat L(\hat\theta)-\log p_0(\hat\theta)+O(1/N)
\end{equation}
where $d$ is the number of parameters (i.e. the dimension of $\theta$), $\hat\theta$ is the maximum likelihood estimate of $\theta$ and $\hat L(\theta)$ is the Hessian matrix of the log-likelihood at $\theta$.

The knowledge that the generative model must be in a family of parametric distribution allows one to project the sample on the manifold spanned by $p(\hat s|\theta)$ and to extract information from $\hat s$, i.e. to estimate $\hat \theta$. This makes it possible to extract information from the sample even when $\hat H[k]=0$ and our analysis would suggest that the sample does not contain any information on the generative model. For example, this is the case when all outcomes are observed only once ($s_i\neq s_j$ for all $i\neq j$). 

%
Yet, in the absence of information on the underlying parametric model, it is reasonable to assume that models that accurately describe the data cannot deliver more information than the information that the sample contains on the generative model, i.e. $D_{KL}\left(p(\theta|\hat s)||p_0(\theta)\right)\le N\hat H[k]$. Considering only the term 
$D_{KL}\left(p(\theta|\hat s)||p_0(\theta)\right)\simeq \frac{d}{2}\log N$, which is the one considered in Bayesian Information Criterium (BIC) model selection \cite{schwarz1978}, this provides an upper bound to the number of parameters that can be estimated from the data
\begin{equation}
d\le\frac{2N\hat H[k]}{\log {N}}.
\end{equation}
This relation suggests that samples with a higher value of the relevance permit to estimate a larger number of parameters, as also advocated in Ref. \cite{HM}.

Eq. (\ref{eq:Dkl}) also shows how the relation between informativeness of samples and criticality manifests in the standard setting of point estimate in statistics. Let us focus on exponential models of the type
\[
p(s|\theta)=\frac{1}{Z}e^{\theta \cdot \phi(s)},
\]
where $\theta=(\theta_1,\ldots,\theta_d)$ is a vector of parameters and $\phi(s)$ is a vector of statistics.
Then, $\hat L$ coincides with the Fisher Information matrix and it has the nature of a susceptibility matrix:
\[
L_{a,b}=\frac{\partial^2}{\partial\theta_a\partial\theta_b}\log Z=\frac{\partial\langle\phi_a\rangle}{\partial\theta_b}.
\]

Eq. (\ref{eq:Dkl}) is the difference between the term $-\log p_0(\hat\theta)$, which quantifies how surprising  $\hat \theta$ would have been {\em a priori}, and the differential entropy of the posterior distribution of  $\theta$, which is a Gaussian variable with mean $\hat \theta$ and covariance $\hat L^{-1}(\hat\theta)/N$. 
Hence, samples can be very informative either because $\hat\theta$ is {\em a priori} very surprising, or because, {\em a posteriori}, the uncertainty on $\theta$ is reduced considerably. The latter occurs if the second term in Eq. (\ref{eq:Dkl}) is as large as possible. Hence, most informative samples are those for which the susceptibility is maximal. 
If the model allows for a ``critical point'' $\theta_c$ at which the susceptibility is very large (and that would diverge in an infinite system), then most informative samples are critical in the sense that $\hat\theta\approx\theta_c$ 
(see also Ref. \cite{Mastromatteo}). 


\section{The Asymptotic Equipartition Property and most informative representations}

In this and the next section, we move from the analysis of a single sample to that of the statistical properties of a system that is encoding a most informative representation. Considering the latter as a statistical mechanical system, we show that statistical criticality is equivalent to an exponential density of energy states (Eq. \ref{ZipfWE}). 

Imagine a data generating process that we think of as draws from an unknown distribution $p(\vec x)$. To fix ideas, $\vec x$ can be thought of as an $n$ dimensional vector with $n \gg 1$ (e.g. a digital picture or gene expression profile of a cell). The different components $x^{(\alpha)}$ of $\vec x$ are dependent random variables and the structure of their dependence is the central object of interest. 
Barring uninteresting cases of strong interactions where the structure of dependencies is easily detectable, we focus on cases where the variability in $\vec x$ is large. More precisely, we assume that the entropy of $\vec x$ is proportional to $n$ and focus on the set of typical $\vec x$ for which
\begin{equation}
\label{AEPx}
-\frac{1}{n} \log p(\vec x) \simeq h_0=-\frac{1}{n}\sum_{\vec x}p(\vec x)\log p(\vec x).
\end{equation}
The Asymptotic Equipartition Property \cite{CoverThomas} ensures that, under the conditions specified above, with probability close to one, all $\vec x$ generated from $p(\vec x)$ satisfies Eq. (\ref{AEPx}). In other words, {\em typically}, all $\vec x$ have the same probability $p(\vec x)\sim e^{-nh_0}$. 

A representation $p(\vec x|s)$ is a mapping from the space of data $\vec x$ to a space $\mathcal{S}$ of labels, so that 
\begin{equation}
\label{}
p(\vec x) = \sum_{s\in\mathcal{S}} p(\vec x|s)p(s).
\end{equation}
Upon defining {\em energy levels}\footnote{Following \cite{Mora,BialekUSSC}, we adopt a statistical mechanics analogy where we refer to the labels $s$ as ``states'' and to the variable $E_s$ as the ``energy'' of state $s$. In information theoretic terms, $E_s$ can be interpreted as a coding cost.} as
\begin{equation}
\label{ }
E_s=-\log p(s),
\end{equation}
the main result of this section is:

\paragraph{Proposition} 
\label{propZipfWE}
{\em For a maximally informative representation, the number $W(E)$ of states $s$ with energy $E_s=E$, is given by}
\begin{equation}
\label{ZipfWE}
W(E)=W_0e^E,
\end{equation}
{\em for all values of $E$ that occur with a non-negligible probability.}
\vskip 0.5cm

A detailed derivation of this result is given in the appendix and it relies on an iterated use of the AEP. First, the AEP is used to characterise the typical points $\vec x$ that we expect will be generated by $p(\vec x)$, and to conclude, as discussed above, that they are equiprobable. Second, we argue that a representation is maximally informative if it assigns the same label $s$ to {\em similar} data points, i.e. to points that differ only by irrelevant details that can be considered as noise. Therefore, the data $\vec x$ conditional on the labels $s$ should have the same properties of a vector of high dimensional independent random variables. This means that $\log p(\vec x|s)\simeq -nh_s$ attains the same value, for all points generated from $ p(\vec x|s)$, with probability close to one\footnote{This is exactly true, with probability one, for maximum entropy models of the type 
\[
p(\vec x)=\frac{1}{Z}e^{\sum_\mu g^\mu\phi^\mu(s(\vec x))},\qquad Z=\sum_{\vec x}e^{\sum_\mu g^\mu\phi^\mu(s(\vec x))}
\]
where $g^\mu$ are parameters and all sufficient statistics $\phi^\mu$ depend on the data $\vec x$ only through $s(\vec x)$. In this case, $p(\vec x|s)=e^{-nh_s}$ is constant, for all $\vec x$ for which $s(\vec x)=s$, with $nh_s$ which is the logarithm of the cardinality of the set $\{\vec x:~s(\vec x)=s\}$.}. In other words, the AEP provides the natural variable (i.e. the log-probability) for efficient representations. Finally, for all values of $E$ which occur with a non-negligible probability (i.e. such that $|\log p(E)|/n\ll 1$), the two results above imply Eq. (\ref{ZipfWE}).

\begin{figure}[ht]
\centering
\includegraphics[width=0.76\textwidth]{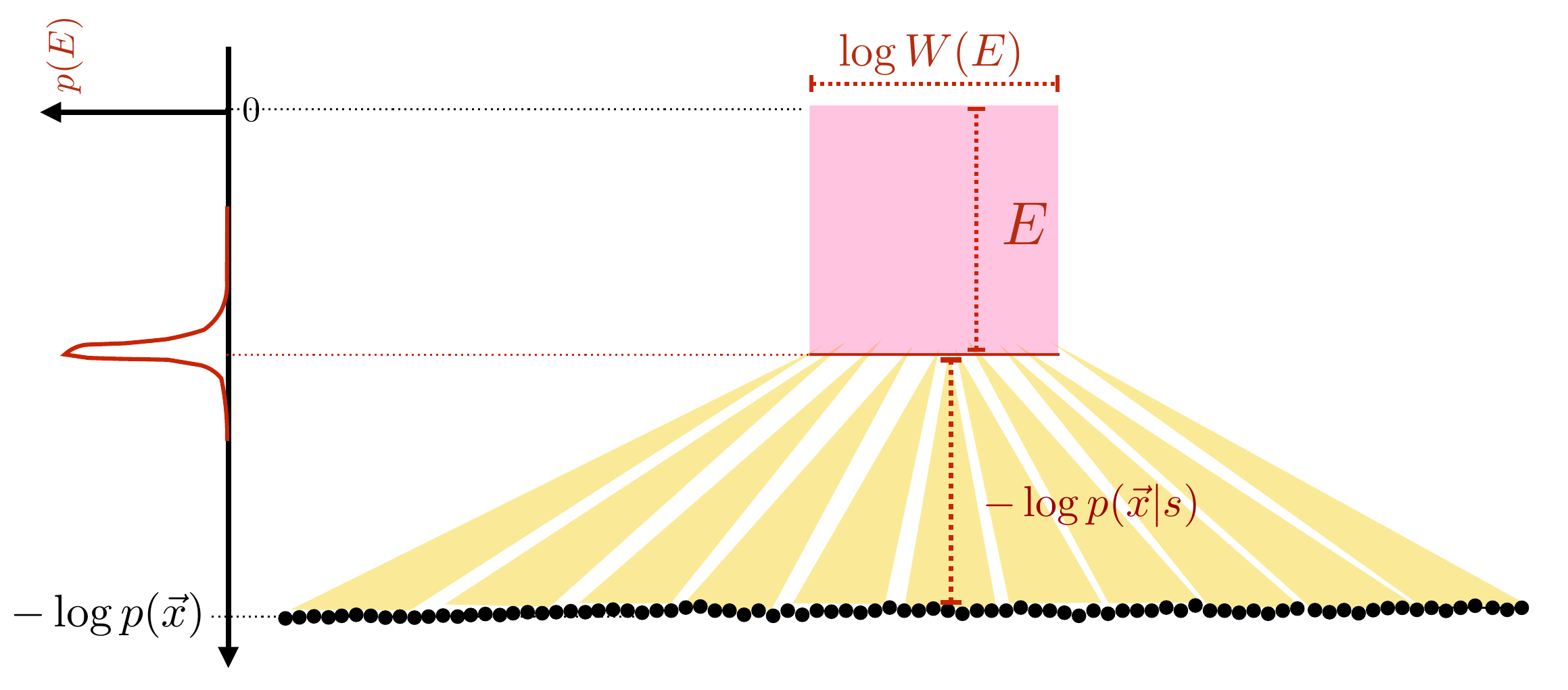}
\includegraphics[width=0.8\textwidth]{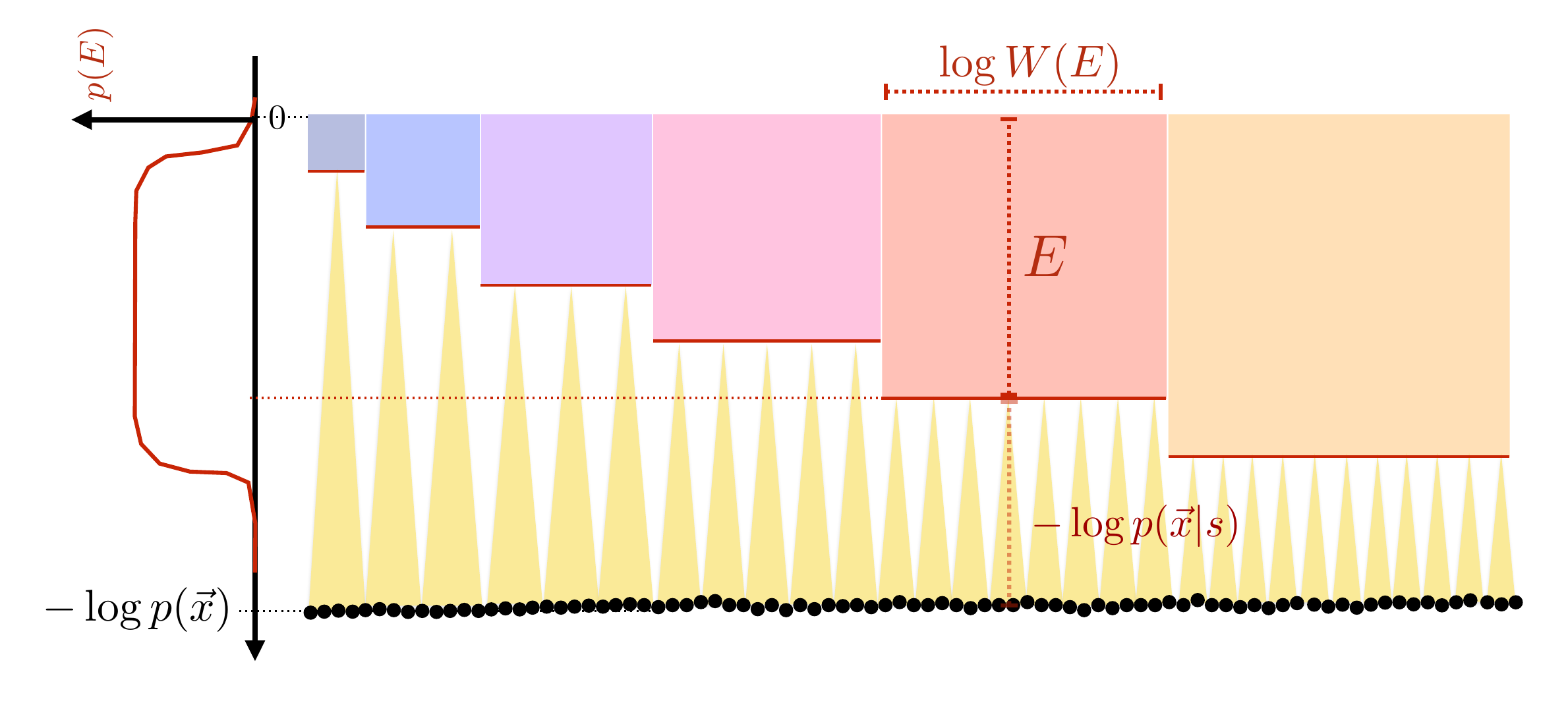}
\caption{\label{FigAEP} Sketch of the argument leading to Eq. (\ref{ZipfWE}). The typical set of points $\vec x$ ($\bullet$) is decomposed in sets of typical draws from $p(\vec x|s)$ (yellow triangles) each corresponding to a different state $s$. The number $W(E)$ of states that correspond to an energy level $E=-\log p(s)$ is related to $E$ be the AEP, Eq. (\ref{ZipfWE}). Structure-less generative processes or totally uninformative representations $s$ correspond to a narrow distribution of energy levels (top). Non-trivial statistical dependencies are revealed by a maximally informative representations, thus resulting in a broad distribution of energies (bottom). When $W(E)=W_0e^{E}$, the distribution of energies $p(E)=W(E)e^{-E}$ becomes flat.}
\end{figure}

Fig. \ref{FigAEP} sketches the idea of the proof. If $\vec x$ is a vector of independent components, then Eq. (\ref{ZipfWE}) holds in a single point, i.e. $E_s\simeq E$ for all $s\in\mathcal{S}$ (see Fig. \ref{FigAEP}, top). This corresponds to a structure-less data generating process $p(\vec x)$. The same holds if $p(\vec x)$ is a data generating process with a rich structure and the variable $s$ is totally uninformative about it\footnote{A simple realisation of this scenario is when $p(\vec x|s)=\delta_{s,s(\vec x)}$, where $s(\vec x)$ is an arbitrary, randomly drawn label in $\mathcal{S}$ assigned to each point $\vec x$ independently.}. Yet, if the data generating process has a rich structure {\em and} $s$ is a maximally informative representation, then $E$ exhibits a broad variation, so that $p(E)=W(E)e^{-E}$ is flat over a broad range of $E$, as indicated in Fig. \ref{FigAEP} (bottom). 
Notice that, interpreting $S(E)=\log W(E)$ as a thermodynamic entropy, as e.g. in Ref. \cite{Mora}, Eq. (\ref{ZipfWE}) corresponds to a linear relation $S(E)=S_0+E$ between energy and entropy, with slope equal to one. 
In other words, maximally informative representations of data with a non-trivial structure, are necessarily characterised by a smooth energy landscape, or a flat free energy landscape $F=E-S\simeq {\rm const}$, over a broad range of values of $E$. 

The linear relation $S(E)=S_0+E$ is equivalent to Zipf's law \cite{Mora}. In order to see this, let us consider a sample  $\hat s$ of $N$ i.i.d. draws from $p(s)$ that corresponds to a maximally informative representation of a dataset $\hat x=(\vec x_1,\ldots,\vec x_N)$. If $\hat s$ samples the energy landscape described above, the number $k_s$ of times that a state $s$ occurs in the sample, is expected to be given by $k_s\simeq Ne^{-E_s}$. The number of states $s$ that occur $k$ times in the sample, which is $km_k$, should match the degeneracy $W(E)=W_0e^E$ for $E\simeq -\log (k/N)$. This implies $m_k\sim k^{-2}$, which is Zipf's law.
In the light of the discussion of previous sections, Eq. (\ref{ZipfWE}) provides a characterisation of maximally informative representations that generate maximally informative samples at the optimal relevance -- resolution trade-off ($\mu=1$).




These results are fully consistent with the observation of Ref. \cite{MSN}, that Zipf's law arises in the presence of hidden variables. In order to see this, it is necessary to revert the logic of the arguments above. Consider $n\gg 1$ independent (or weakly dependent) variables $\vec x=(x_1,\ldots,x_n)$ drawn independently from the same probability $p(x|h)$ that depends on a variable $h$. Under these conditions, $\vec x$ satisfies the Asymptotic Equipartition Property (AEP) \cite{CoverThomas}. This states that there is a typical set $\mathcal{A}_h$
such that (asymptotically as $n\to\infty$) {\em i)} all $\vec x$ in $\mathcal{A}_h$ have the same probability
\[
-\frac{1}{n}\log p(\vec x|h)=-\frac{1}{n}\sum_{i=1}^n\log p(x_i|h)\simeq u(h)=-\sum_x p(x|h)\log p(x|h)
\] 
and {\em ii)} almost surely, $\vec x$, which is drawn from $p(\vec x|h)$, belongs to $\mathcal{A}_h$, and (because of this) {\em iii)} the number of typical $\vec x$'s is equal to $|\mathcal{A}_h|\approx e^{n u(h)}$. 
If one defines the entropy $\sigma(h)$ as the logarithm of this number divided by $n$, then one has that $\sigma(h)=\frac{1}{n}\log|\mathcal{A}_h|\approx u(h)$. This is equivalent to Zipf's law \cite{Mora,MSN}, provided $u(h)$ varies over a broad range of values. Hence, a necessary condition for obtaining Zipf's law is that the hidden variable $h$ induces a variation of $u(h)$ over a wide range. In this case, the probability (density) to observe a value $nu$ of $-\log p(\vec x|h)$ is given by
\begin{equation}
\label{flatE}
p(u) = \int dh p(h)e^{n[\sigma(h)-u(h)]}\delta(u-u(h)).
\end{equation}
Since $\sigma(h)\approx u(h)$ by the AEP, the distribution $p(u)$ remains broad and it does not concentrate. 
Ref. \cite{MSN} corroborates this argument with convincing numerical experiments for few cases. Ref. \cite{ACL} also points out that a uniform distribution in Eq. (\ref{flatE}) is equivalent to Zipf's law.
Although the variable $u$ and the variable $E$ introduced above do not coincide (see the appendix), Eq. (\ref{flatE}) is equivalent to a flat distribution in $E$. 
In the light of the main result of this section, we identify the hidden features of Ref. \cite{MSN}  as those providing a maximally informative representation. 

\section{The thermodynamics of efficient representations}

The discussion above provides a general derivation of efficient representations with a given entropy $H[s]$ that, in analogy with the discussion of Section \ref{maxinfosamples}, we shall call resolution. As we have seen in the previous section, efficient representations correspond to maximally informative samples with $\mu=1$. In this section, we analyse how this result generalises for different values of $H[s]$.

Again, we consider a high dimensional vector of inputs $\vec x$ which is generated from an unknown distribution $p(\vec x)$. 
Let us focus on a situation where the dependence structure in the inputs $\vec x$ is very rich. We set out to seek a representation in terms of discrete states $s(\vec x)$ that captures this dependence. The generative model induces a distribution 
\begin{equation}
p(s)=\sum_{\vec x:~s(\vec x)=s}p(\vec x)
\end{equation}
on the set of states. 
As before, let us define {\em energy levels} $E_s=-\log p(s)$ and assume that these take values in a discrete set $E_s\in\mathcal{E}$. Notice that energy levels are bounded in a finite interval $E_s\in [0,E_{\max}]$ because $0<p(s)<1$ for all $s$\footnote{In particular the most unlikely state is expected to be exponentially unlikely in the dimensionality $n$ of the inputs. Hence, we expect that $E_{\max}$ is proportional to $n$.}. Let $W(E)$ be the number of energy levels with $E_s=E$. $W(E)$ is an integer which is expected to be exponentially large in the dimensionality $n$ of the inputs. The properties of the representation depend on the degeneracy $W(E)$ of energy levels. Hence, our goal is to find the $W(E)$ that correspond to most informative representations.

The distribution of the random variable $E$ is given by
\begin{equation}
\label{PE}
P(E)=W(E)e^{-E},\qquad \sum_{E\in\mathcal{E}}W(E)e^{-E}=1.
\end{equation}
Notice that the average energy $\langle E\rangle=H[s]$ corresponds to the entropy of the labels, whereas the entropy of $E$ is given by
\begin{equation}
\label{HEeq}
H[E]=H[s]-H[s|E],\qquad H[s|E]=\langle\log W(E)\rangle.
\end{equation}
Here, $H[s|E]$ provides a measure of the noise \cite{SMJ}, which arises from the residual degeneracy between states that cannot be distinguished. Indeed, as shown in the previous section, points generated from $p(\vec x|E_{s(\vec x)}=E)$ cannot be distinguished from random vectors. Eq. (\ref{HEeq}) provides the same decomposition as in Eq. (\ref{decamp}) of the information content of a point $s$ in terms of noise and {\em useful} information. This is why we shall use the term relevance both for $\hat H[k]$ (referred to a sample) and for $H[E]$ (at the population level). Therefore:

\paragraph{Proposition} {\em Maximally informative representations $s(\vec x)$, at a given $H[s]$, are those with a degeneracy $W(E)$ of states such that the noise $H[s|E]$ is minimal, or equivalently for which the relevance $H[E]$ is maximal.}

\vskip 0.5cm

In order to find optimal representations, we introduce a Lagrange multiplier $\nu$ enforcing the constraint on $H[s]$ and we maximise $H[E]+\nu H[s]$ on $W(E)$. This leads to
\begin{equation}
\label{WEeff}
W(E)=W_0e^{(1+\nu)E},\qquad E\in[0,E_{\max}],
\end{equation}
where $W_0$ ensures that the normalisation in Eq. (\ref{PE}) is satisfied. 
Notice that this corresponds to a linear dependence between the entropy $S(E)=\log W(E)=(1+\nu)E+\log W_0$ and the energy $E$. 

\begin{figure}[ht]
\centering
\includegraphics[width=0.8\textwidth]{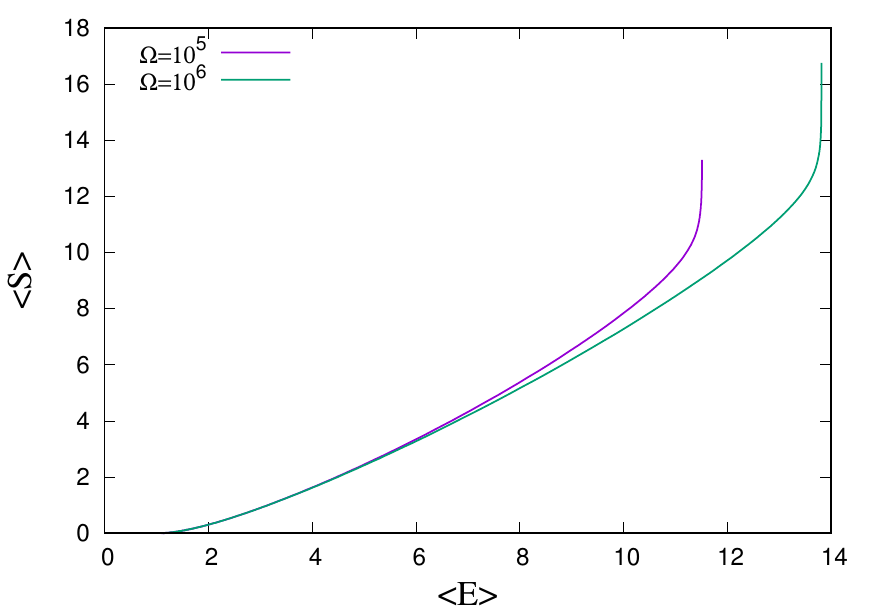}
\caption{\label{fig:thermanal} Relation between entropy $\langle S\rangle=H[s|E]$ and energy $\langle E\rangle=H[s]$ for maximally informative representations with $\Omega=10^5$ and $10^6$ states. These curves are obtained from the minimisation of $\langle S\rangle$ at fixed $\langle E\rangle$, while also fixing the number of possible states $\Omega=\sum_{E\in\mathcal{E}}W(E)$, through a Lagrange multiplier, in the optimisation over $W(E)$.}
\end{figure}

The behaviour of the expected value of the entropy $S(E)$ versus the expected value of the energy $E$ in most informative representations is shown in Fig. \ref{fig:thermanal}. The convexity of this curve is unconventional in statistical mechanics, where the entropy is a concave function of the energy. This unconventionality is a consequence of the fact that, while statistical mechanics seeks the maximal entropy distribution $p(s)$ at a fixed degeneracy $W(E)$ of energy levels, most informative representations are characterised by a degeneracy $W(E)$ of $E$ that minimises the average entropy $\langle S\rangle=H[s|E]$ at fixed $p(s)=e^{-E_s}$. In spite of the fact that the energy constraint is the same, the associated Lagrange multipliers have very different meanings. Indeed, $\nu$ cannot be thought of as an inverse temperature\footnote{Small (large) values of $\nu$ correspond to low (high) energies $\langle E\rangle$.}. Rather, $\nu$ in maximally informative representations has a natural interpretation in terms of the resolution-relevance trade-off. Indeed, the slope of the curve in Fig. \ref{fig:thermanal} is given by $\nu+1$, which means that if the resolution $H[s]$ is reduced by one bit, the noise $H[s|E]$ decreases by $1+\nu$ bits. Eq. (\ref{HEeq}) then implies that $H[E]$ increases by $\nu$ bits. The region $\nu>0$ corresponds to ``redundant'' representations, whereas for $\nu<0$, some informative bits are ``lost in compression''. These two regions are separated by the point $\nu=0$ for which $H[E]$ is maximal.

As a consequence of the maximisation of entropy in statistical mechanics, the distribution of energies is sharply peaked, hence $H[E]\simeq 0$, which is at the basis of the equivalence between the micro-canonical and the canonical ensembles\footnote{In the micro-canonical ensemble, the entropy is given by Boltzmann's formula $\log W(E)$ or by its average $H[s|E]$. In the canonical ensemble, it is given by the Gibbs-Shannon entropy $H[s]$. By Eq. (\ref{HEeq}) these are approximately the same when $H[E]\approx 0$. As an example, in the Ising ferromagnet in $d$ dimensions, the number of energy levels is proportional to the number $n$ of spins, hence $H[E]\le\log n$. Both $H[s]$ and $H[s|E]$ instead are proportional to $n$.}. Given that equilibrium statistical mechanics and most efficient representations arise from opposite optimisation principles, it is hardly surprising that statistical criticality is so rare in the former (and it generally requires parameter fine tuning) and ubiquitous in the latter, as we shall see.

Maximally informative representations correspond to partitions $s(\vec x)$ of the space of inputs such that the typical distribution of observed energy levels $E_s$ is broad. These broad distributions in the energy levels correspond to wide flat minima in the free energy landscape. The width of these minima can be measured by the variance $C=\langle(E-\langle E\rangle)^2\rangle$ of the energy levels, that can be computed\footnote{The standard trick of taking the second derivative of $\log \langle e^{-\beta E}\rangle$ can be used to obtain $C$ in Eq. (\ref{Cv}).} using Eq. (\ref{WEeff})
\begin{equation}
\label{Cv}
C=E_{\max}^2f(\nu E_{\max}),\qquad f(z)=\frac{1}{z^2}+\frac{1}{2-2\cosh(z)}.
\end{equation}
For $E_{\max}\gg 1$, the variance has a sharp maximum at $\nu=0$ of width $1/E_{\max}$, because $f(z)\simeq 1/12 -z^2/240+\ldots$ for $z\ll 1$. This reflects the fact that, at $\nu=0$, the distribution of energies is as wide as possible. At this particular point, the energy spectrum is used as efficiently as possible (see Ref. \cite{SMJ}). When $\nu>0$, high energy states overweigh low energy ones, whereas when $\nu<0$, the distribution of energy levels is skewed on low energy states. 

Even though the generative model $p(\vec x)$ is unknown, it is possible to see how the optimality of a representation $s(\vec x)$ manifests in a typical sample of $N$ independent draws from $p(\vec x)$. Also, while the energy levels are unknown, the expected number of sampled points with energy levels $E_s$ in an interval around $E_k=-\log(k/N)$ should be proportional to $W(E)e^{-E}dE$ and it should match the number $k m_kdk$ of states observed in the corresponding interval  of the frequency $k$. From this, 
\[
km_k\sim \left.W(E)e^{-E}\left|\frac{dE}{dk}\right|\right|_{E=-\log(k/N)}\sim \frac{W(E_k)e^{-E_k}}{k}
\]
For an efficient representation that satisfies Eq. (\ref{WEeff}), this implies that $m_k\sim k^{-2-\nu}$, which is the same result derived in Eq. (\ref{powlaw}) with $\mu=1+\nu$. Therefore, the maximisation of $H[E]$ at fixed $H[s]$, that underlies Eq. (\ref{WEeff}), is equivalent to the maximisation of $\hat H[k]$ at fixed $\hat H[s]$ within a sample. This justifies the use of the term relevance for both $\hat H[k]$ and $H[E]$.

Notice that when no constraint is imposed on the resolution (i.e., $\nu=0$) one recovers Zipf's law. More compressed efficient representations correspond to broader frequency distributions ($\nu<0$) whereas less compressed ones give rise to steeper frequency distributions ($\nu>0$). 

The fluctuation of the energies $E_k=-\log(k/N)$ can be computed within a finite sample. The above discussion implies that fluctuations should be maximal for representations with $\nu=0$, i.e. that satisfy Zipf's law. The authors of Refs. \cite{Mora,retina,BialekUSSC} have advocated a thermodynamic construction based on a single sample, considering a modified distribution $p_\beta(s)=\frac{1}{Z(\beta)} k_s^{\beta}$. The second derivative of $\log Z(\beta)$ wrt $\beta$ yields the variance of $E_s=-\log (k_s/N)$ over $p_\beta$, that can be interpreted as a specific heat $C(\beta)$ in this analogy. When the underlying distribution satisfies Zipf's law, one finds a maximum of $C(\beta)$ at $\beta=1$ \cite{Mora,retina}, which is consistent with the picture discussed above. 
At the maximum, we expect $C(\beta_{\max})\simeq E_{\max}^2/{12}$. Ref. \cite{retina} measured $C(\beta)$ in the neural activity of a population of neurons and found that  $C(\beta_{\max})$ increases with the number $n$ of neurons recorded. 
For most informative samples, this measure reveals how $E_{\max}$ increases with the dimensionality $n$ of the input $\vec x$.

\section{Relation with the Information Bottleneck method}

Let us consider a generic task in unsupervised learning. Data $\vec x$ is produced from an unknown data generating process $q$ and we wish to extract a representation $s$ of the data points that can shed light on $q$. For example, in a data clustering task, the data $\hat x$ consists of a sequence of $N$ different objects $\vec x_i$. The task is that of grouping these points into classes, by attaching a label $s_i=s(\vec x_i)$ to each point, that may highlight features of the generating process $q$, such as similarities and differences among data points. Also, $\vec x$ can represent patterns that a deep neural network aims at learning and $s$ the state of one of the layers in the architecture \cite{SMJ}. 
Viewed as a Markov chain, this corresponds to
\begin{equation}
\label{eq:markovVSH}
q \rightarrow \vec x \rightarrow s.
\end{equation}
A formal approach to the task consists in looking for the association $p(s|\vec x)$ that solves the problem
\begin{equation}
\label{IB}
\max_{p(s|\vec x)}\left[I(q,s)-\beta I(s,\vec x)\right],
\end{equation}
where the first term of the optimisation function is the information that $s$ contains on the generative process and the second term penalises redundant representations. Eq. (\ref{IB}) is very close to the Information Bottleneck (IB) method \cite{IB} in spirit. The main difference is that in the IB, $q$ and $\vec x$ are given, so IB deals with the {\em supervised} learning task of determining the representation $s$ that encodes the relation between $q$ and $\vec x$ in an optimal manner. Here, $q$ is unknown and we are interested in the {\em unsupervised} task of learning an optimal representation of the data. As long as $q$ is unknown, Eq. (\ref{IB}) remains a formal restatement of the problem\footnote{For application of IB to unsupervised learning problems, such as geometric data clustering, see e.g. \cite{IBclustering}.}. 

However, if we restrict to a finite sample, $q$ can be replaced with the empirical frequency $\hat p_s=k_s/N$, thus making the problem in Eq. (\ref{IB}) well defined. The solution to this problem yields representations for which $\hat s$ is a maximally informative sample. In order to show this, note that if we replace $q$ with the empirical frequency $\hat p_s=k_s/N$, the first term in Eq. (\ref{IB}) becomes 
\begin{equation}
\label{IB1}
I(q,s)=I(k,s)=\hat H[k]-\hat H[k|s]=\hat H[k]
\end{equation}
where the last equality derives from the fact that $k_s$ is a function of $s$. The second term is minimal when $s$ is a function of $\vec x$, so that $I(s,\vec x)= H[s]- H[s|\vec x]= H[s]$ can be replaced by the entropy of $s$. We note, in passing, that  Ref.  \cite{DIB} shows that, conversely, if $I(s,\vec x)$ is replaced by $H[s]$ in Eq. (\ref{IB}) then $p(s|\vec x)$ reduces to a deterministic mapping $s(\vec x)$. 

In a finite sample, the second term in Eq. (\ref{IB}) is given by the resolution $I(s,\vec x)= \hat H[s]$. Taken together with Eq. (\ref{IB1}), this turns Eq. (\ref{IB}) into the constrained maximisation of $\hat H[k]$ subject to a constraint on $\hat H[s]$, which is what defines maximally informative samples.

The substitution $q\to\hat p$ amounts to the statement that conditional on $s$, $\vec x$ contains no information on $q$. Indeed, $I(\vec x,\hat p|s)=0$ because $\hat p$ is a function of $s$. This is equivalent to reversing the Markov chain in Eq. (\ref{eq:markovVSH}) as $q \rightarrow s \rightarrow \vec x$\footnote{This argument parallels the definition of sufficient statistics \cite{CoverThomas}: When  $q=f(\cdot|\theta)$, then, the Markov chain in Eq. (\ref{eq:markovVSH}) reads as $\theta \rightarrow \vec x \rightarrow s$. If $s$ is chosen as a sufficient statistic for $\theta$ then, conditional on $s$, the data $\vec x$ do not contain any information on $\theta$. This implies that the chain can be reversed, i.e. $\theta \rightarrow s \rightarrow \vec x$.}, so that $q$ becomes the generative model of $s$. 

In summary, maximally informative samples are the solution of an optimisation problem similar to IB, with the important difference that while IB is a {\em supervised} learning scheme, maximally informative samples are the outcome of an {\em unsupervised} learning task. Indeed, the IB addresses the issue of maximally compressing an input $\vec x$ to transmit relevant information that reconstructs a given output $q$~\cite{shwartz2017}, whereas the definition of maximally informative samples takes the frequency $k$ of the internal representations $s$ as output features.
This highlights the fact that relevance is defined with respect to a pre-specified output in the IB, whereas the approach discussed here quantifies relevance with respect to an internal criteria. We remark, in this respect, that the first term $I(k,s)=\hat H[k]$ in the optimisation function does not depend at all on the relation between $s$ and $\vec x$, but only on the distribution of the former. Note also that, in contrast to the rate-distortion curves typical of IB where relevance $I(q,s)$ is an increasing function of channel capacity $I(s,\vec x)$, here the relation is not monotonic. This is consistent with the findings of Ref. \cite{IBfinitesize}, that finite size effects generate a similar bending in the IB curves.

\section{Conclusions}

The first aim of this paper is to clarify the derivation and nature of the relevance $\hat H[k]$, recently introduced in \cite{HM,MMR}, as a measure of the useful information that a sample contains on the generative model. We do this by relating our approach to the standard approach employed in parametric statistics. As a byproduct, we also derive an estimate of the maximal number of parameters that can be estimated from a dataset, in the absence of prior knowledge on the generative model.
Furthermore, we characterise the properties of maximally informative samples and the trade-off they embody between resolution and relevance. This offers a different explanation of the widespread occurrence of statistical criticality \cite{statcrit}.  Our results suggest that any complex interacting system of many degrees of freedom, when expressed in terms of relevant variables -- those embodying a maximally informative representation at the resolution afforded by a finite sample -- should exhibit statistical criticality. 
In particular, we find that Zipf's law characterises the statistics of maximally informative samples at the optimal trade-off between resolution and relevance. The principle of maximal relevance suggests {\em why} statistical criticality may emerge, independently of any self-organisation or parameter fine-tuning mechanism \cite{SOC}. Different mechanisms may be required to explain {\em how} this principle is implemented in specific systems.

The second aim of this paper is to characterise the statistical properties of systems that optimally encode the dependence structure of high dimensional data.  We find that, within a statistical mechanics description, 
maximally informative representations are characterised by an exponential energy density of states (Eq. \ref{WEeff}). This feature emerges from a principle of maximal relevance, which is conjugate to the maximum entropy principle in statistical mechanics. 
In the light of these results, it is not surprising that hidden layers' representations extracted by deep neural networks exhibit broad distributions, as observed in \cite{SMJ}. In particular, the frequency of observed states of the hidden layer with optimal generation ability follows Zipf's law very accurately \cite{SMJ}. Within Restricted Boltzmann Machines, Ref. \cite{Hennig} finds that statistical criticality emerges as a consequence of the fact that the information content of the encoded inputs (the variable called $E_s$ here) acts as a hidden variable. 
Some of us have confirmed that optimal coding within Minimum Description Length theory, also operates very close to the limit of maximal relevance \cite{MDL}. 
It is suggestive to relate the wide and flat energy landscape in the space of inputs, implied by the AEP, for most informative representations to the presence of wide and flat energy minima in the space of weights that has been suggested  \cite{Zecchina} to be at the origin of the impressive performance of deep learning. 

A flat energy landscape and broad frequency distributions are expected to emerge in general in all systems that are designed to extract efficient representations\footnote{The notion of efficiency that is implied here is defined in terms of the information that the representation carries on the generative model of the states of the environment. Loosely speaking, maximally informative representations are optimal generative models of the states of the environment.}. This extends, as argued in \cite{Hidalgo}, to living adaptive and evolutionary systems, both at the individual and at the collective level. In line with Ref. \cite{Hidalgo}, it is suggestive to think of the principle of maximal relevance as a distinctive feature of living systems, whose activity depends on the efficiency of their internal representation of the environments they live in \cite{TkacikBialek2014}. This principle distinguishes living systems from physical systems, which are instead subject to the principle of maximal entropy of statistical mechanics. In this perspective, statistical criticality in living systems \cite{Mora} would arise as a signature of this distinctive feature. 

Besides its appeal as a simple rationale for the occurrence of broad distributions and Zipf's law in many domains
 \cite{Zipf,Immune,Mora5405,retina}, we believe our results uncover a very general principle underlying maximally informative representations. As such, we expect that it will show its most useful application as a guideline to evince useful information from high dimensional data and/or for extracting efficient representations.
   
 \section*{Acknowledgements}
 We acknowledge interesting discussions with M. Abbott, E. Aurell, J. Barbier, R. Monasson, T. Mora, I. Nemenman, N. Tishby and R. Zecchina. This research was supported by the Kavli Foundation and the Centre of Excellence scheme of the Research Council of Norway (Centre for Neural Computation) (RJC and YR), by the Basic Science Research Program through the National Research Foundation of Korea (NRF), funded by the Ministry of Education (2016R1D1A1B03932264) (JJ), and, in part, by the ICTP through the OEA-AC-98 (JS).
    
 \appendix
 \section{Derivation of Eq. (\ref{ZipfWE})}
 
The aim of this section is to prove Eq. (\ref{ZipfWE}), which provides a characterisation of a maximally informative representation of a generic data generating process. For concreteness, let us assume that a data point is an $n$-dimensional vector\footnote{For simplicity, we assume the components $x_a$ are drawn from a finite set $\chi_a$, so that we can refer to the AEP in its basic form \cite{CoverThomas}. } $\vec x=(x_1,\ldots,x_n)$, with $n\gg1$, and that the generating process can be represented as a probability distribution $p(\vec x)$, from which $\vec x$ is drawn. We also assume that $p(\vec x)$ satisfies the Asymptotic Equipartition Property (AEP). This states that, for a small $\epsilon>0$, almost surely, all points $\vec x$ generated from $p(\vec x)$ belong to the typical set
\begin{equation}
\label{empty1}
\mathcal{A}=\left\{\vec x:~\left|-\frac{1}{n}\log p(\vec x)-h_0\right|<\epsilon\right\},\qquad h_0=-\frac{1}{n}\sum_{\vec x}p(\vec x)\log p(\vec x).
\end{equation}
As a consequence, the number of typical points is $|\mathcal{A}|\simeq e^{nh_0}$. 
In words, this ensures that, with very high probability, all $\vec x$ have the same probability $p(\vec x)\sim e^{-nh_0}$. This is equivalent to assuming that all points in a finite sample are equally likely. 
 
Still, $p(\vec x)$ contains non-trivial statistical structure. In order to capture this statistical structure, we introduce a variable $s\in\mathcal{S}$, in such a way that, conditional to $s$, the vector $\vec x$ can be considered as noise. This implies that (almost) all $\vec x$ drawn from $p(\vec x|s)$ satisfy
\begin{equation}
\label{empty2}
-\frac{1}{n}\log p(\vec x|s)\simeq h_s,\qquad h_s=-\frac{1}{n}\sum_{\vec x}p(\vec x|s)\log p(\vec x|s).
\end{equation}
Put differently, if we define $s$-typical sets 
\begin{equation}
\label{empty3}
\mathcal{A}_{s}  =\left\{\vec x:~\left|-\frac{1}{n}\log p(\vec x|s)-h_s\right|<\epsilon\right\},
\end{equation}
the AEP ensures that almost all $\vec x$ drawn from $p(\vec x|s)$ belong to $\mathcal{A}_s$. Two points $\vec x,\vec x'\in \mathcal{A}_s$ can be considered similar, since they differ only by irrelevant details (e.g. noise). The set $\mathcal{A}$ can be decomposed  as 

\begin{equation}
\label{empty4}
\mathcal{A} = \bigcup_{s\in\mathcal{S}} \mathcal{A}_{s}
\end{equation}
into $s$-typical sets $\mathcal{A}_s$. The AEP also implies that the number of $s$-typical points is $|\mathcal{A}_s|\sim e^{n h_s}$, because all $s$-typical points have the same probability $p(\vec x|s)\simeq e^{-nh_s}$ and $\vec x$ drawn from $p(\vec x|s)$ almost surely belong to $\mathcal{A}_s$. Therefore, since 
\begin{equation}
\label{empty5}
p(s)=\sum_{\vec x\in \mathcal{A}_s}p(\vec x)\approx\frac{| \mathcal{A}_s|}{| \mathcal{A}|}
\end{equation}
then
\begin{equation}
\label{eqx}
E_s\equiv -\log p(s)\simeq n(h_0-h_s),
\end{equation}
in the sense that $E_s/n\to h_s-h_0$ when $n\to\infty$. Now, let us consider the set of $E$-typical points
\begin{equation}
\label{empty6}
\mathcal{A}_E=\bigcup_{s:E_s=E} \mathcal{A}_s.
\end{equation}
For all $\vec x\in \mathcal{A}_E$ we have
\begin{eqnarray}
-\frac{1}{n}\log p(\vec x|E) & = & -\frac{1}{n}\log \sum_{s:E_s=E}p(\vec x|s)p(s|E)\\
 & \simeq & -\frac{E}{n}+h_0+\frac{1}{n}\log W(E)\label{eqcco}
\end{eqnarray}
where we have used the fact that, for all $\vec x \in \mathcal{A}_E$, the sum above is dominated by only one value of $s$, for which $p(\vec
x | s) \simeq e^{-nh_s}\simeq e^{-nh_0+E}$ by equation (\ref{eqx}). For all other values of $s' \neq s$, $p(\vec x | s) \approx 0$. Now we observe that, for all $\vec x\in \mathcal{A}_E$
\begin{eqnarray}
h_0 & \simeq & -\frac{1}{n}\log p(\vec x) \\
 & = &  -\frac{1}{n}\log p(\vec x|E)-\frac{1}{n}\log p(E)\label{eqqq}\\
 & \simeq &  -\frac{1}{n}\log p(\vec x|E)
\end{eqnarray}
for all values of $E$ for which $p(E)$ is not exponentially small, i.e. for which 
$(\log p(E))/n\to 0$ as $n\to\infty$. 
This, along with Eq. (\ref{eqcco}), implies that $W(E)\simeq e^E$, which is Eq. (\ref{ZipfWE}).
 
\bibliographystyle{unsrt}
\bibliography{EfficientRepresentationsIB.bib}

\end{document}